\documentstyle[editedvolume]{crckapb}

\def\apj#1{{\em Astrophys. J.} {\bf #1}}
\def\mn#1{{\em Mon. Not. R. astr. Soc.} {\bf #1}}
\def\aa#1{{\em Astron. Astrophys.} {\bf #1}}
\def\aas#1{{\em Astron. Astrophys. Suppl.} {\bf #1}}

\def\nat#1{{\em Nature,} {\bf #1}}
\def\apjs#1{{\em Astrophys. J. Suppl.} {\bf #1}}

\def\be{\begin{equation}}
\def\ee{\end{equation}}
\def\bea{\begin{eqnarray}}
\def\eea{\end{eqnarray}}
\def\etal{{\it et al.}\ }
\def\Mpc{$h_{100}^{-1}$~{\rm Mpc}}

\begin{opening}

\title
{REGULARITY OF THE LARGE-SCALE STRUCTURE OF THE UNIVERSE}

\author{J. EINASTO}

\institute{Tartu Observatory, EE-2444 T\~oravere, Estonia}

\end{opening}

\runningtitle{REGULARITY OF THE STRUCTURE}

\begin{document}

\section{Introduction}

The observed structure of the Universe is hierarchical.  Galaxies and
clusters of galaxies are concentrated within elongated filamentary
chains of various richness. High-density regions of the Universe form
superclusters consisting of one or several clusters of galaxies and
chains of galaxies surrounding and joining clusters.  The space
between filaments is void of galaxies.  Superclusters and voids form a
continuous network of alternating high- and low-density regions in the
Universe.

Systems of galaxies are formed by density waves of wavelength which
correspond to the size of system. It is believed that density waves
have a Gaussian distribution; on small scales the distribution of
galaxy systems is actually well represented by a random process.
There exists, however, growing evidence that on larger scales the
distribution may have some regularity. In particular, the
supercluster-void network shows the presence of a non-random character
in the distribution.  In this talk I shall discuss this evidence and
its possible explanation.

\section{Evidence for the regular location of high-density regions}

\begin{figure}
\vspace{6.3cm}
\caption{Distribution of Abell-ACO clusters in high-density regions in
supergalactic coordinates. Only clusters in superclusters with at
least 8 members are plotted.  The supergalactic $Y=0$ plane coincides
almost exactly with the Galactic equatorial plane and marks the
Galactic zone of avoidance. In the left panel some superclusters
overlap due to projection, in the right panel only clusters in the
southern Galactic hemisphere are plotted and the projection effect is
small.
}
\includegraphics{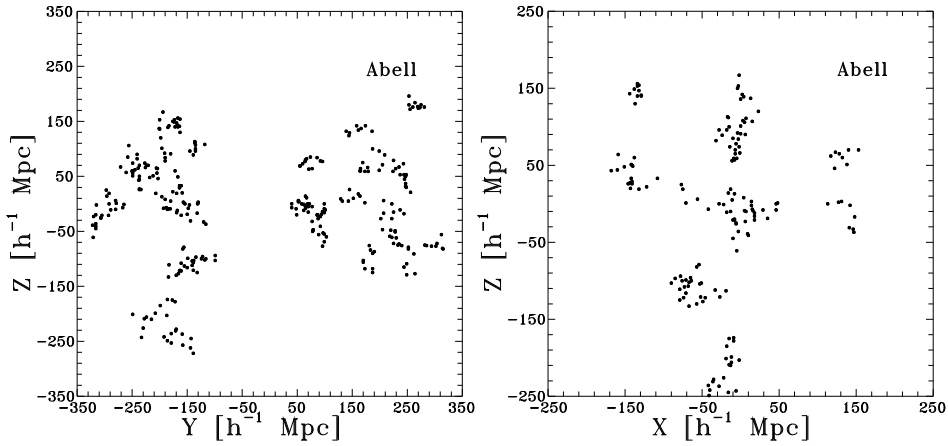}
\label{fig1}
\end{figure}

Already first studies of the distribution of superclusters and voids
have shown that large cluster-defined voids are surrounded by
superclusters and resemble cells with low-density interiors and
high-density environs.  Examples are the Northern Local Void
surrounded by the Local, Coma and Hercules superclusters, and the
Bootes void located between the Hercules, Bootes and several other
superclusters \cite{je78,k81,l95}.  Cluster defined voids have a
rather large size, of the order of 100~\Mpc\ \cite{zes82}. A similar
scale was found from the study of the cluster correlation function
which has a secondary maximum near 130~\Mpc\ due to the concentration
of clusters in superclusters on the other side of large voids
\cite{kopylov,mo,eg93,fetisova}. However, these early studies did not
ask the question: Has the supercluster-void network some large-scale
regularity?

The first clear evidence for the existence of a regularity in the
distribution of galaxies on large scales came from the pencil beams
around the northern and southern Galactic pole~\cite{beks}. It was
demonstrated that high- and low-density regions in the distribution of
galaxies alternate with a rather constant step of 128~\Mpc. Nearest
peaks in the distribution of galaxies coincide in position and
redshift with superclusters \cite{bahcall}, thus one may conclude that
the regularity should be a property of high-density regions in the
Universe in general.  This discovery raised many questions. The basic
counter-argument was that on the basis of an one-dimensional survey one
cannot draw conclusions on the distribution of matter in three
dimensions.

To investigate the global regularity of the supercluster-void network
our group in Tartu used Abell-ACO clusters of galaxies
\cite{abell,aco}. This is the deepest all-sky survey available
presently. To understand the nature of the distribution we used first
cosmographic approach, and plotted clusters of galaxies in
superclusters of various richness to show their dependence on the
large-scale environment. Plots of Abell-ACO clusters of galaxies
located in very rich superclusters with at least 8 member-clusters are
shown in Figure~1 \cite{me94,me95,me97}.  We see a quasi-regular
network of superclusters and voids.  High-density regions are
separated from each other by roughly constant intervals of $\approx
120$~\Mpc.

\section{The correlation function and the power spectrum of clusters 
of galaxies}

The correlation function of a quasi-regular network of clusters of
galaxies should have periodic maxima and minima which correspond to
mutual distances of high- and low-density regions. This phenomenon is
actually observed; see Figure~2 (Einasto \etal 1997a). The power
spectrum of clusters of galaxies is also shown in Figure~2
\cite{einasto}, it was determined from the correlation function using
the Fourier transform. The power spectrum is peaked on wavelength
$\lambda_0=120$~\Mpc\ which corresponds to the size of the step of the
supercluster-void network.  On short wavelengths the spectrum can be
approximated by a power law with index $n=-1.8$.  On long wavelengths
the spectrum is not well determined, within observational errors it is
compatible with the Harrison-Zeldovich spectrum which has power index
$n=1$.

\begin{figure}
\vspace{4.3cm}
\caption{ The correlation function (left panel) and the power spectrum
(right panel) for clusters of galaxies. Dashed lines show the error
corridor. The solid line in the right panel shows the power spectrum
of the standard CDM model ($h=0.5$, $\Omega=1$). The amplitude of the
cluster power spectrum was decreased by a bias factor $b^2=3^2$. }
\includegraphics{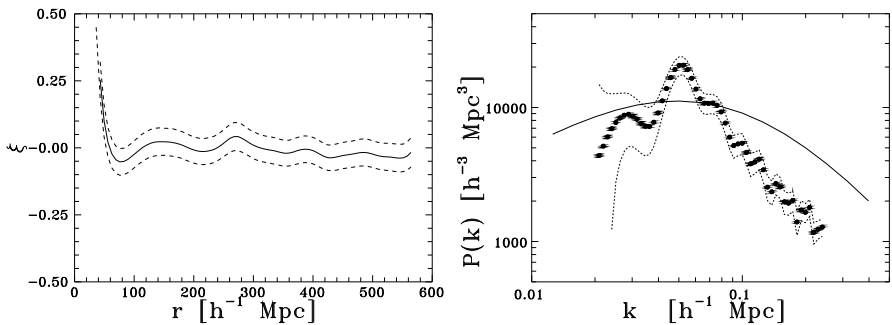}
\label{fig2}
\end{figure}

Recent investigations of the power spectrum of galaxies and clusters
of galaxies demonstrate similar behaviour of the spectrum on large
scales.  Peacock (1997) and Gaztanaga and Baugh (1997) determined the
three-dimensional power spectrum from the projected distribution of
APM galaxies; Retzlaff \etal (1997) used Abell-ACO clusters of
galaxies but a completely independent method of data analysis; Tadros
\etal (1997) calculated the power spectrum for APM clusters. These
four independent studies show that the power spectrum on scales
shorter than $\approx 120$ \Mpc\ is approximately a power law with
index between $n=-1.8$ and $n=-1.9$; the spectrum has a rather sharp
maximum near the wavelength 120 \Mpc, and approaches the
Harrison-Zeldovich regime on longer wavelengths. The exact position of
the maximum and the shape of the spectrum near the maximum vary
between studies, but these variations are well in the limits expected
from errors of the determination of the power spectrum.

\section{Comparison with CMB data}

The present matter power spectrum is generated by two different
processes: by the inflation of the Universe, and by the hot radiation
dominated era before recombination. Inflation determines the initial
power spectrum of matter. During the hot phase density fluctuations
within the horizon are damped, thus on small wavelengths the amplitude
of the spectrum decreases; this behaviour is described by the transfer
function. The transfer function is given by cosmological parameters:
the density of matter in various populations (baryon, cold and hot
dark matter), the Hubble constant etc. To estimate the possible role
of these processes Atrio-Barandela \etal (1997) performed calculations
of the angular spectrum of temperature anisotropy of the cosmic
microwave background (CMB).  The temperature anisotropy spectrum has a
maximum around wavenumber $l\approx 200$ due to acoustic oscillations
of the hot plasma before recombination. The shape of the angular
spectrum around this maximum is very sensitive to the initial spectrum
and cosmological parameters, and can be used as a test.

\begin{figure}
\vspace{6cm}
\caption{
  Comparison of matter power spectra and radiation temperature
  anisotropy spectra with cluster and CMB data.  Dots with $1\sigma$
  error bars give the observations: the measured cluster spectrum
  (Einasto \etal 1997a) in the left panel and the Saskatoon data on
  CMB temperature anisotropies (Netterfield \etal 1997) in the right
  panel. The scale-free model spectra (short-dashed lines) were
  computed using the following parameters: $h=0.6$, $\Omega_{b}=0.07$,
  $\Omega_{c}=0.23$, and $\Omega_{\Lambda}=0.7$.  The cluster spectrum
  based model (solid lines) was calculated using $h=0.6$,
  $\Omega_{b}=0.08$, $\Omega_{c}=0.92$, and $\Omega_{\Lambda}=0$; the
  double power approximation of the peaked spectrum (long-dashed
  lines) using $h=0.6$, $\Omega_{b}=0.05$, $\Omega_{c}=0.95$, and
  $\Omega_{\Lambda}=0$.  To compare matter power spectra and observed
  cluster spectrum we used a bias factor $b_{cl}\approx 3$ for
  clusters.}
\includegraphics{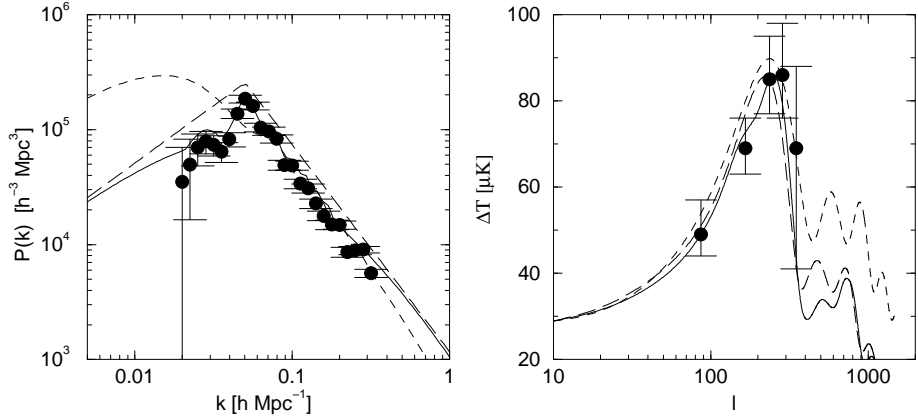}
\end{figure}

Calculations were performed for the standard CDM-type model with a
scale-free initial spectrum (which has a constant power index
$n\approx 1$), and two models with the peaked power spectrum.  For a
set of cosmological parameters, Atrio-Barandela \etal find the matter
transfer function, and the matter and radiation power spectra.  It is
assumed that the Universe has a flat geometry. For the calculations
the CMBFAST package was used \cite{cmbfast2}. To estimate the goodness
of a fit of a particular set of cosmological parameters authors
calculate  parameter $\chi^2$ comparing model data with observations
made in Saskatoon \cite{saskatoon}.

Recent measurements favour a low density of matter in the Universe,
but an open Universe seems to be excluded as in this case the first
acoustic maximum of CMB angular spectrum shifts to too high
frequencies. Thus a CDM model with large cosmological constant is of
particular interest.  In a model with large cosmological constant the
baryon density is comparable to the density of the dark matter, and
acoustic oscillations of the hot plasma before recombination have an
enhanced amplitude~\cite{szalay}. This model can be considered as a
candidate for the peaked power spectrum.

Calculations by Atrio-Barandela \etal show that, using an appropriate
set of cosmological parameters, temperature anisotropy spectra of
different models are very similar in the range of multipoles observed
in the Saskatoon data (see Figure~3).  In other words, the present CMB
data are not sufficient to discriminate between models.  The matter
power spectra are also similar on short wavelengths, but on medium and
long scales they are different. The scale-free model with a large
cosmological constant has a broad maximum at large wavenumber
($k\approx 0.01~h_{100}~{\rm Mpc^{-1}}$); the maximum of the first
acoustic oscillation occurs at $k\approx 0.1~h_{100}~{\rm Mpc^{-1}}$,
and is of rather small relative amplitude.  Both scales are outside
the allowed range of the position of the spike in the cluster
spectrum: $k_{0}=0.052 \pm 0.005~h_{100}~{\rm Mpc^{-1}}$
\cite{einasto}.  Thus the observed spike is not related to acoustic
oscillations in the baryon--photon plasma as assumed by Szalay (1997)
but must have a different origin.

\section{Initial spectrum}

\begin{figure}
\vspace{6cm}
\caption{
Initial power spectra. The left panel shows initial power spectra for
models, calculated for parameters listed in the caption of Figure~3.
The right panel gives a theoretical initial power spectrum suggested
by Starobinsky; the dashed line shows for comparison the
standard scale-free initial power spectrum. }
\includegraphics{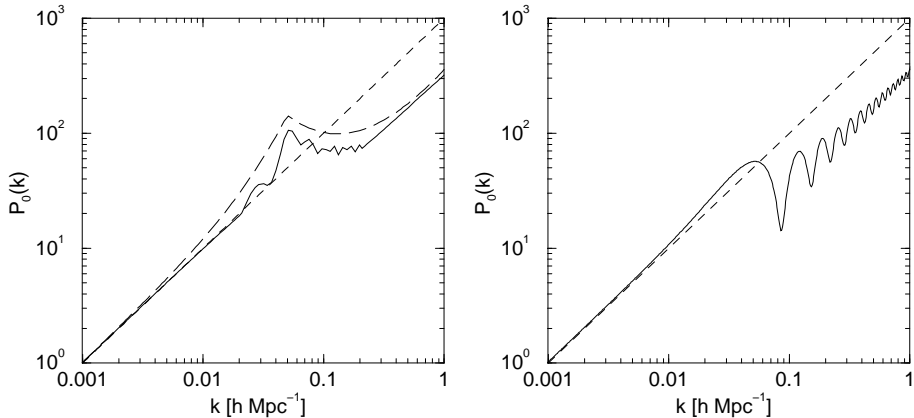}
\end{figure}

The initial power spectra for our models are shown in Figure~4. We see
that the presence of a peak in the present power spectrum leads to a
peak also in the initial spectrum. Another important deviation of the
initial spectrum calculated from the observed spectrum is the presence
of a break: the amplitude of the spectrum on shorter wavelengths is
smaller than that expected for a single scale-free spectrum.  The
initial spectrum found from the observed power spectrum has three
parameters: the position and relative height of the peak, and the
relative amplitude of the break in respect to the scale-free spectrum.

It is interesting to note that an initial power spectrum similar to
those found from the observed cluster spectrum was suggested by
Starobinsky (1992). Starobinsky assumed that the inflation of the
Universe may have had a more complicated form than usually accepted.
If the inflaton field was evolving through a non-smooth point of its
potential then a change in the first derivative of the potential
generates a spike followed by a break in the initial spectrum. Such an
initial spectrum is shown in the right panel of Figure~4.

\section{Evolution of the structure}

To investigate the influence of various regions of the power spectrum
to the structure of the Universe we performed N-body calculations of
several simple models with a constant power index $n\approx -1.5$ on
small scales (wavenumber $k\geq k_0$), a maximum at the wavenumber
$k=k_0$, and various shape and power indices at longer wavelengths
\cite{eg93,f95}.

This test shows that the fine structure of superclusters is entirely
given by density waves shorter than the maximum of the spectrum. The
scale of the supercluster-void network is fixed by the wavelength of
the maximum of the spectrum, $\lambda_0= {2\pi/k_0}$. Long waves
modulate the structure: if there is no power on longer scales then all
supercluster have practically equal masses and are regularly
distributed.  The variance of supercluster masses and relative
distances increases if one raises the amplitude of the spectrum at long
wavelengths.  The shape of the spectrum around the maximum determines
the regularity of the supercluster-void network. If the transition
from a positive spectral index at long wavelengths (Harrison-Zeldovich
spectrum) to a negative index at small scales is sudden then the
network is fairly regular. On the other hand, if the transition is
smooth then the network is irregular.

Density perturbations of medium and small scale located at the top of
density waves having the largest amplitude give rise to the formation
of superclusters of galaxies. Here the overall amplitude of
perturbations is the largest, respective small-scale systems (clusters
and groups of galaxies) have large masses, they form first and evolve
more rapidly.  On the contrary, similar medium and small-scale
perturbations located near the minima of large-scale density waves
have a low overall amplitude, respective systems of galaxies have
small masses, and their formation starts later. Thus our simple models
explain the hierarchy of galaxy systems and show why large
cluster-defined voids are not empty but contain galaxy systems similar
to superclusters but with much lower masses \cite{zes82,l95}. These
simple models also show the importance of the study of the
distribution of superclusters -- they are determined by the behaviour
of the power spectrum near its maximum. In this region differences
between various cosmological models are the largest.

\section{Discussion}

The comparison of optical and CMB spectra has shown that no
combination of cosmological parameters can reproduce the spike at
$k=k_{0}$: the existence of a broad maximum is an intrinsic property
of all scale-free models.  A universe with a smooth maximum of the
matter power spectrum and randomly distributed phases of density
fluctuations will have a random distribution of high- and low-density
regions \cite{einasto2,einasto1}. Such a distribution is in
contradiction with the distribution of superclusters.  Thus present
data favour the interpretation of the observed spectrum in terms of a
peaked, broken scale initial spectrum.

Present data are, however, not accurate enough to make a final
decision on the initial power spectrum of matter. One point of concern
is the power spectrum itself: Is the power spectrum at all an
appropriate statistic to give a full description of the distribution
of matter in the Universe? As Szalay (1997) has shown, a change of
phases for two distributions having identical power spectra may change
the appearance dramatically. A simple toy model with a regular
cellular structure after phase scrambling transforms to a completely
random sample. As the Universe on scales of the supercluster-void
network definitely has some regularity, then phase information seems
to be essential to describe the structure. So far no appropriate
statistic has been suggested.

Another point of concern has been the use of Abell-ACO clusters as
indicators of the large-scale structure. Several authors have argued
that this catalogue may contain systematic errors, and the use of a
machine-selected catalogue of clusters as the APM catalogue is to be
preferred. To check this point we have compared the distribution of
Abell-ACO and APM clusters of galaxies in rich superclusters.  This
analysis shows that rich superclusters found on the basis of these
completely independent cluster samples are very similar. Quantitative
tests lead to similar conclusions \cite{me94}. On the other hand, the
volume covered by the APM sample is much smaller than the volume
covered by the Abell-ACO sample. Following J\~oeveer and Einasto
(1978) and Zeldovich \etal (1982) we call large low-density regions
surrounded by superclusters as cells.  The APM sample covers just 1 --
2 such cells only in the southern Galactic hemisphere whereas the
Abell-ACO sample contains information on a dozen of cells in both
hemispheres.  Thus, to study the large-scale distribution of
high-density regions, the Abell-ACO sample is much better suited.

The comparison of CMB and matter spectra shows that definite
differences between models with scale-free and peaked initial power
spectra are expected for small angular scales in the CMB spectrum, and
on very large scales in the present matter spectrum. Future
experiments, both for optical and CMB regions, are concentrated on
these wavelengths. Thus the choice between the models will be much
easier in the near future.

Whatever the answers to these questions are, one can say that 
the distribution of high-density regions in the Universe brings us
information on the very early stages of its evolution.

\vspace{0.2cm}

I thank H. Andernach, F. Atrio-Barandela, M. Einasto, S.  Gottl\"ober,
V.  M\"uller, A. Starobinsky, E. Tago and D. Tucker for fruitful
collaboration and permission to use our joint results in this review
talk, and A. Szalay and D. Wood for discussion.  This study was
supported by the Estonian Science Foundation.

\end{document}